# Nanopores – a Versatile Tool to Study Protein Dynamics

Sonja Schmid, Cees Dekker

*Department of Bionanoscience, Kavli Institute of Nanoscience, Delft University of Technology,
Van der Maasweg 9, 2629 HZ Delft, The Netherlands.*
Correspondence: s.schmid@tudelft.nl

**ABSTRACT**

Proteins are the active working horses in our body. These biomolecules perform all vital cellular functions from DNA replication and general biosynthesis to metabolic signaling and environmental sensing. While static 3D structures are now readily available, observing the functional cycle of proteins – involving conformational changes and interactions – remains very challenging, e.g., due to ensemble averaging. However, time-resolved information is crucial to gain a mechanistic understanding of protein function. Single-molecule techniques such as FRET and force spectroscopies provide answers but can be limited by the required labelling, a narrow time bandwidth, and more. Here, we describe electrical nanopore detection as a tool for probing protein dynamics. With a time bandwidth ranging from microseconds to hours, it covers an exceptionally wide range of timescales that is very relevant for protein function. First, we discuss the working principle of label-free nanopore experiments, various pore designs, instrumentation, and the characteristics of nanopore signals. In the second part, we review a few nanopore experiments that solved research questions in protein science, and we compare nanopores to other single-molecule techniques. We hope to make electrical nanopore sensing more accessible to the biochemical community, and to inspire new creative solutions to resolve a variety of protein dynamics – one molecule at a time.

# INTRODUCTION

Proteins are the working horses in our body [1]. They convert energy into function, and form complex dynamic protein-protein interaction networks. Protein function relies crucially on structured domains and structural flexibility [2] that are both encoded in the peptide chain and its post-translational modifications (PTMs; see glossary). Also natively unfolded parts that may obtain a specific 3D arrangement only upon binding to their interaction partners play an important role [3]. The protein world offers an inexhaustible wealth of active and passive dynamic nanoscale phenomena ranging from the extreme precision during DNA replication [4] (polymerases, helicases), mechanical force generation (myosin [5], flagellar motor [6], dynein, kinesin [7], proteasome [8]), to other energy conversion (ATPsynthase [9], ion pumps [10], bacterio-rhodopsin [11], light harvesting complexes [12]), as well as sensing and signalling (von Willebrandt factor [13], tip-link cadherins [14,15], piezo protein [16,17], kinases, isomerases, countless PTM regulators [18–21]). These are just a few examples of the wide variety of protein functions that rely critically on conformational dynamics and protein-protein interactions. However, this key aspect – the functional dynamics that a single protein performs – is challenging to observe experimentally. Established structural biology techniques such as X-ray diffraction or cryo-electron microscopy [22] have outstanding spatial resolution but are blind for dynamic effects that are happening in solution, while NMR [23], EPR [24], SAXS [25], and SANS [26,27] suffer from ensemble averaging.

Fortunately, various single-molecule techniques have taken on the challenge to resolve protein dynamics in real time at room-temperature: most prominently, single-molecule fluorescence techniques (e.g., smFRET [28,29]), single-molecule force spectroscopies (e.g. AFM [30], magnetic [31,32] or optical [33,34] tweezers), and – more recently – electrical nanopore detection [35]. None of these techniques can compete with the 3D Angstrøm resolution of cryoEM or x-ray crystallography. But in exchange for some spatial resolution, they resolve dynamic processes performed by just one molecule or one functional entity – and notably in real-time at room temperature and in solution. The fundamental advantage of time-resolved single-molecule approaches is that different functional states can be distinguished in a molecular ensemble, and that following one molecule over time allows one to resolve the timescales involved, to quantify kinetic rate constants, and ultimately to uncover the energetic driving forces that control protein function. In short, time-resolved single-molecule techniques provide a direct view on how proteins perform function, which constitutes the essential ingredient to move from static protein structures to understanding dynamic protein function at the nanoscale.

**The scope of this review**

So why, as a protein fan, should you care about nanopores? This is the central question in this review. Well, you should care specifically if you are interested in protein dynamics, i.e., in pushing the boundaries beyond static structural biology. The unrivalled time range accessible within one nanopore experiment is perfectly suitable to study protein dynamics, in most cases even label-free (as reviewed below). Moreover, nanopores have already proven their utility in many DNA and RNA sequencing applications performed outside traditional nanopore-centric labs [36–40], and portable handheld nanopore sequencing devices have been realized and marketed [41]. Building on the ongoing success of nanopore technology in DNA/RNA sequencing [42,43], we review here the next step: nanopore solutions to research questions in protein science.

We target this review at scientists who are interested in original quantitative biochemistry, and protein function in particular, while we simultaneously hope to stimulate enthusiasm for proteins among nanopore experts. In the first part of this brief review, we introduce in a nutshell the basic concept of nanopores. We describe typical experiments, various pore designs, the characteristics of the electrical nanopore signals - all from a protein sensing perspective. In the second part, we highlight a few creative nanopore applications that contributed to our understanding of protein function, and we compare electrical nanopore sensing to other single molecule techniques. Clearly, by zooming in on a few selected recent findings on proteins, we pass over many seminal contributions that have shaped the nanopore field. We refer the inclined reader to excellent reviews on the general rise of nanopore technology [44–47], and the chronology from first electrophysiology experiments to DNA sequencing [48] and its commercialization.

## WHAT ARE NANOPORES?

A nanopore is a most simple and elegant sensor: it literally consists of a hole in a membrane that can sense molecules in solution one by one, even label-free (**Figure 1A**). The insulating membrane separates two compartments filled with an electrolyte (i.e., a conducting buffer solution). Depending on the experiment, the pore diameter can range from sub-nanometer size to several tens of nanometers. When a voltage is applied between both compartments, an ionic current flows through the nanopore, which can be measured with an amplifier. In the most basic sensing scheme, an analyte – such as a single protein – reaches the pore (by diffusion or additional driving forces described below), where it blocks the flow of ions leading to a characteristic resistive pulse with a current blockade $\Delta I$, and an event duration $\Delta t$ that signals the residence time of the protein in the pore (**Figure 1B**). The magnitude of the ionic current blockage depends on the size (molecular weight) and even on the shape of the analyte [49]. The event duration varies greatly for proteins, depending on the specific solid-state or biological pore structure, possible interactions with the pore walls, and the dominating driving force to the nanopore: while 'near-ballistic' translocation events happen within microseconds [50], specific trapping conditions offer much longer observation times (**Fig 1C**). The development of such new long-term sensing schemes was a major step forward, making nanopores more interesting for protein science than ever. We describe several specific examples in the second part of this review, to illustrate how nanopores catch protein functional features that are inaccessible to other methods.

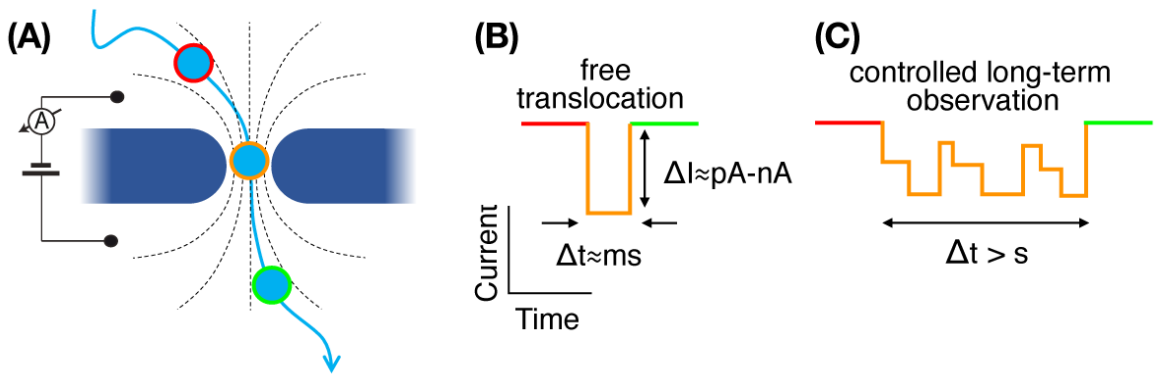

*Figure 1: Nanopores in a nutshell. (A) Voltage is applied across an insulating membrane (dark blue) with a nanopore immersed in buffer, causing a measurable ionic current through the pore. A particle (light blue) approaches the nanopore (red position), partially blocks it (orange position), and subsequently leaves the pore again (green position). Electric field lines are illustrated as dashed lines. (B, C) Schematic nanopore current signals. The position color code of (A) is used to refer to the particle trajectory: a baseline current (red, green) is observed before and after the particle induced current blockade (orange). (B) A resistive pulse caused by free particle translocation provides only a short observation time (orange). (C) Creative experiment designs achieve much longer observation times (orange).*

**The nanopore zoo**

Two classes of nanopores are typically distinguished: solid-state nanopores made by nanofabrication, and self-assembled biological nanopores. Solid-state nanopores have the advantage that they can be made at will, with a wide range of sizes and material properties. Glass nanopipets are the most affordable version of solid-state nanopores (**Figure 2A, left**): a pipet puller is used to produce a glass pipet that narrows down to a small nanopore at its end. While nanopipets have been used in various insightful experiments, their application is limited by their inherent asymmetry and the set material. Cleanroom-fabricated solid-state chips (**Fig. 2A, center**) offer more design flexibility. They often use a ~5x5x0.3 mm silicon chip as a substrate for a much thinner free-standing silicon nitride membrane (several nm to hundreds of nm thick), which is symmetrically accessible. Standard chips with thinned-down silicon nitride membranes are commercially available [51]. Recently, their high-frequency noise and thus signal-to-noise performance was significantly improved by adopting glass instead of silicon as the main substrate material, offering better dielectric and capacitive properties and hence lower noise [35]. Pores can be drilled by diverse techniques, each with their pros and cons: e.g., by controlled dielectric breakdown [52–56], by laser etching [56], using an electron beam in an SEM [57], a focused ion beam in a helium ion microscope [58,59], or a TEM [60] – listed in order of increasing cost, but also in approximate order of control and capacity. Beyond silicon nitride, various other membrane materials have been investigated, including atomically thin 2D materials [61], such as graphene [62], hexagonal boron nitride [63], or molybdenum disulfide [64] (displayed **in Fig 2a, right**). Semiconducting 2D materials have been investigated regarding additional detection strategies, such as sensing in-plane currents through the 2D membrane [65]. In addition, 2D materials attracted great attention regarding

DNA sequencing, because the atomic thickness at the nanopore should provide the ultimate spatial precision for reads along a translocating ssDNA strand. Until today, however, these delicate high-end solid-state sensors suffer from low reproducibility, added noise sources, and prohibitive fabrication cost – all still preventing DNA sequencing applications for the end user. Instead, protein nanopores won the race for nanopore-based DNA sequencing [66,67], admittedly by playing several clever tricks, including additional helper proteins, as described in part 2 of this review.

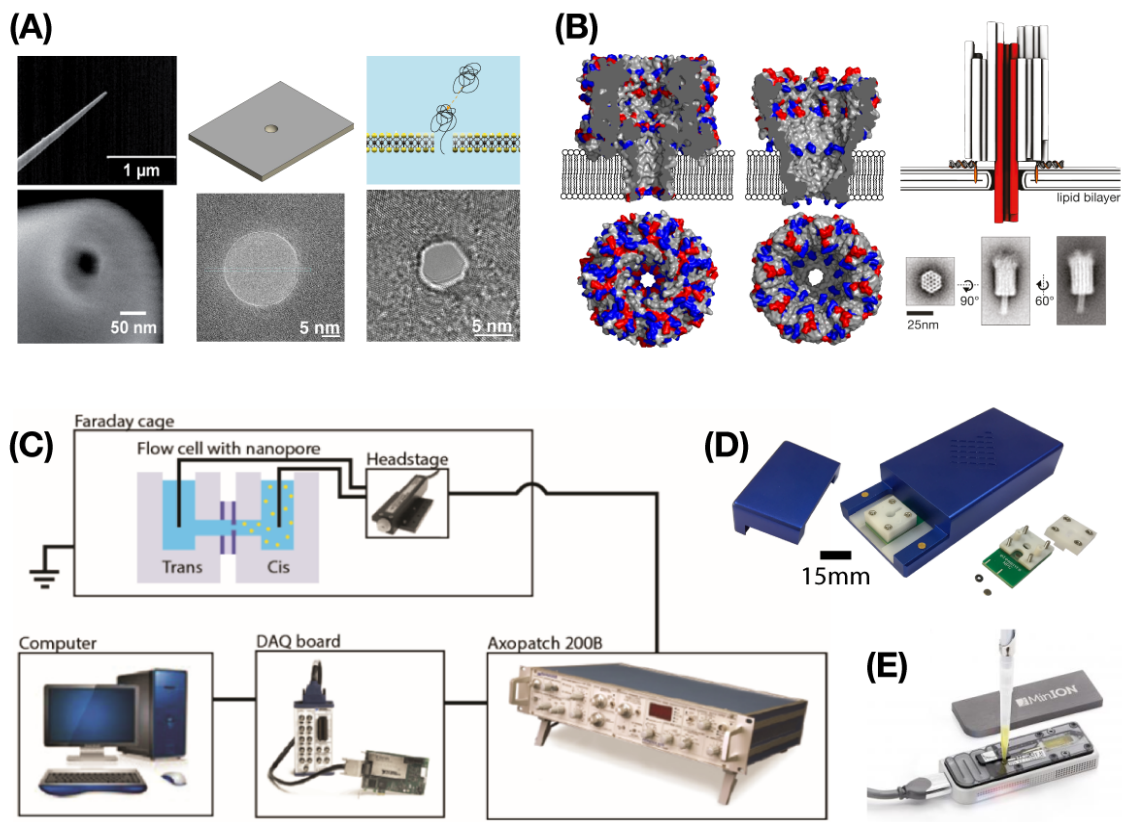

*Figure 2: Various nanopores and instrumentation. (A)* Three prominent examples of solid-state nanopores: (from left to right) glass pipet pulled to yield a sub-50nm pore; glass chips with a TEM-drilled 20nm pore in a silicon nitride membrane; nanopore in the 2D-material molybdenum disulfide. *(B)* Three representative biological nanopores – two protein pores (cut-open view in top panels, and top view in bottom panels; colors encode the charge distribution at neutral pH: positive, blue; negative, red; neutral, gray): the toxin α-hemolysin from Staphylococcus aureus (pdb: 3ANZ [68]); the commonly used M2 mutant of MspA from Mycobacterium smegmatis (pdb: 1UUN [69], mutated); an early synthetic DNA-origami pore with cartoon and TEM images. *(C)* A standard laboratory nanopore setup, with two-compartment flowcell, amplifier headstage, Axopatch 200B amplifier (Molecular Devices), DAQ board digitizer (National Instruments), computer for control & recording. *(D)* More compact integrated instrumentation: the Nanopore Reader (elements). *(E)* The portable, and highly parallelized Minion DNA sequencer (Oxford Nanopores) with 512 channels. Figure sources: panel (A) left [70], center [71], right [72]; panel (B) right [73]; panel (C) [74]; panel (D) adapted from [75]; panel (E) [76].

Biological nanopores [77] differ from their solid-state brothers, in two fundamental ways: (i) a lipid bilayer [78,79] or a block-copolymer membrane [80] serves as the insulating barrier between the buffer compartments, and (ii) the pores come with fixed self-assembled 3D structures and sizes, providing advantageous atomic precision and reproducibility. The α-hemolysin pore shown in **Figure 2B left** is the most famous and widely studied pore protein. However, many other pores have been described [81,82], ranging in size from ~1nm to ~10nm in diameter (e.g. CsgG [83] and PlyAB [84], respectively). In many cases, these protein pores were engineered (with truncations, fusions, point mutations) to serve a specific sensing task [85]. For example, the charges in the pore mouth of MspA were mutated to facilitate the translocation of highly negatively charged ssDNA strands: **Figure 2B, center** shows the charge distribution of the widely used M2-MspA mutant [86]. In addition, entirely synthetic nanopores have been rationally designed using DNA-origami technology (**Fig. 2B, right**). Lipid membrane insertion of these highly charged, ion permeable structures is achieved by attaching hydrophobic anchor molecules, such as cholesterols or porphyrins that spontaneously insert into lipid bilayers [87]. DNA-origami has also been used to control membrane insertion of peptide assemblies [88]. And most recently, a synthetic proteinaceous potassium channel was demonstrated, indicating that protein nanopores with 'custom-made' specifications may become available in the future [89]. Two (manageable) bottlenecks in biological nanopore experiments are the stability of the fragile membrane, and the requirement to insert precisely one pore for single-molecule reads. Hybrid strategies combining solid-state scaffolds for lipid bilayers and biological nanopore sensors have been proposed as an improvement, but classical free-standing lipid bilayer approaches [78,79] still prevail in the literature.

The driving force by which the analyte reaches the pore depends on the analyte's net charge, as well as on pore properties. It is often dominated by electrophoresis (i.e., essentially electrostatic forces acting on the analyte's net charge, which is reduced by charge screening in solution) as in the case of highly negatively charged nucleic acids. However, depending on the nanopore itself, electro-osmosis can instead be the main driving force, especially for less charged analytes such as proteins. Electro-osmosis is a phenomenon arising at charged pore walls, when the applied voltage moves the screening counter-ions uni-directionally along the field lines, dragging water molecules along, and thus causing substantial hydrodynamic flows [84,90]. Such electro-osmosis is very useful for protein sensing, as it allows one to drive proteins to the nanopore regardless of their net charge.

Instrumentation for nanopore experiments is commercially available and affordable. The general experimental setup is similar for solid-state and biological nanopores (**Fig. 2C).** In both cases, an insulating membrane separates two buffer compartments. A few tens to hundreds of millivolts are usually applied across the pore using chloridized silver electrodes, resulting in typical currents in the range of 10 pA to several nA, depending on the pore and the conductivity of the buffer (the electrolyte) used. This current signal is amplified by a low-noise patch clamp amplifier, digitized, and controlled and recorded through a computer. This all-electrical sensing scheme lends itself to miniaturization, which has led to very compact [51] (**Fig 2D**) and highly parallelized handheld devices (**Fig 2E**). Furthermore, megahertz amplifiers have been realized using on-chip architectures [91,92].

**Characteristics of nanopore signals**

Protein-induced nanopore signals arise, because the protein reduces the density of the mobile charge carriers (ions) that occupy the nanopore, leading to a characteristic reduction in electrical conductance. The sensitivity of nanopores for the translocation of proteins and even for their shape and orientation was experimentally demonstrated by various groups [93,94]. Mayer and coworkers showed that near-spherical proteins like streptavidin create uni-modal current blockades, while disk-like immunoglobulin G caused a broad blockade distribution, related to the orientation of the protein in the pore (**Fig 3A**). A different perspective on nanopore sensitivity was given by Bayley and coworkers who showed that human and bovine thrombins are readily distinguished, despite 86% sequence identity [95].

High signals and low noise are needed to resolve protein kinetics. We recently compared the signal-to-noise ratio and individual noise sources in solid-state and biological nanopores [35], and hence we give here only a brief example of how noise scales with time resolution. This is important for every technique, since the detection of protein kinetics requires both a high enough signal-to-noise ratio and a wide enough time bandwidth (defined by the shortest and longest resolvable time interval) [96]. The noise level depends on the chosen low-pass filter frequency and thus time resolution, which is exemplified in **Figure 3B**. The signal-to-noise ratio furthermore scales with the electrolyte conductivity, i.e., the ionic strength of the buffer. As we will see in part 2, already at physiological to moderate salt concentrations that are suitable for proteins, nanopores provide ample SNR to detect protein function.

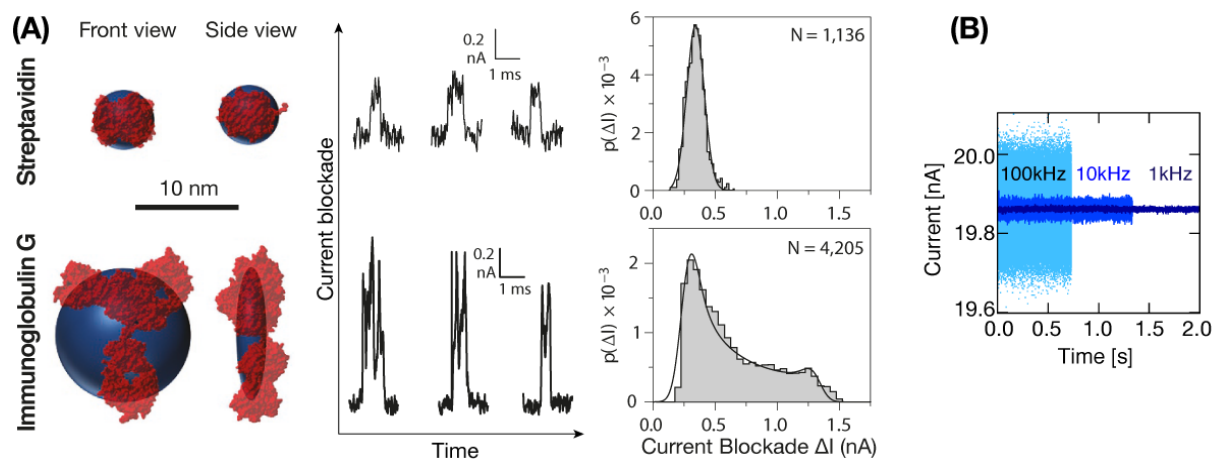

*Figure 3: Nanopore signal characteristics. (A) Current blockades are protein size and shape dependent. Left: front and side view of streptavidin and Immunoglobulin G. Center: protein translocation events for streptavidin (top) and IgG (bottom). Right: corresponding event histograms. All adapted from Ref. [93]. (B) The noise dependence on low-pass filter frequency: current snippets of a 30nm silicon nitride pore in 1M KCl low-pass filtered at 100, 10, 1kHz as indicated.*

# WATCHING PROTEINS AT WORK USING NANOPORES

In this second part of the review, we showcase a few ground-breaking nanopore applications to protein science. We start with nanopore enzymology, the label-free real-time observation of protein enzymatic function.

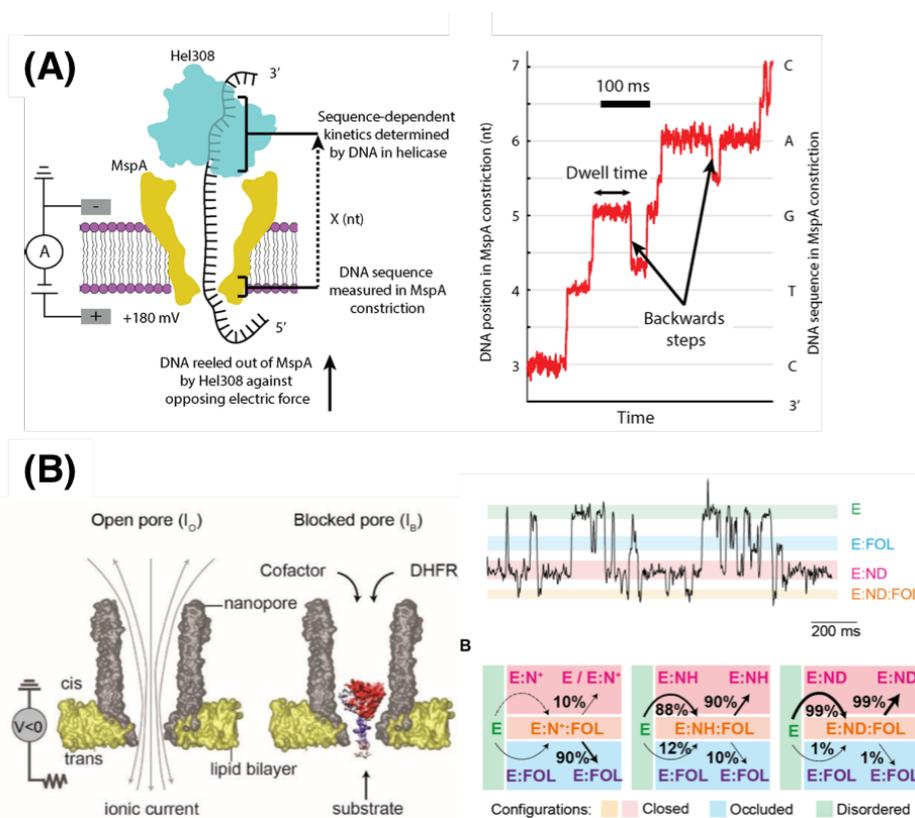

*Figure 4: Live recording of protein function with nanopores. A)* DNA processing by the helicase Hel 308 (cyan), detected using the MspA protein pore (yellow) as steps in the recorded nanopore current [97]. *B)* The functional cycle of dihydrofolate reductase (DHFR) directly observed by electro-osmotic trapping in the ClyA protein pore sensor resolves five functional states (four displayed) [98].

**Nanopore enzymology**

The Akeson and Gundlach labs developed a beautiful experiment that directly resolves ATP-dependent DNA processing by a motor protein, such as a polymerase [67,99] or a helicase [100], thereby achieving the central breakthrough to today's nanopore DNA sequencing. **Fig 4A** shows how the DNA's negative charge is exploited to trap the protein-DNA complex in the MspA pore protein under a positive voltage. In an ATP-dependent way, the helicase Hel308 reels in the ssDNA strand against the electrostatic pulling force (downwards in **Fig. 4A**). In this way, the helicase directly facilitates a slow DNA translocation through the protein nanopore sensor, leaving enough observation time to resolve half-basepair steps at millisecond time resolution, and notably all label-free. While the actual current signal at any given time is affected by approximately four neighbouring nucleotides along the DNA strand, individual base calling can still be achieved by post-hoc signal processing and pattern recognition algorithms. This combination of an ATP-driven motor protein plus a protein nanopore sensor has become the basis for today's commercialized nanopore DNA or RNA sequencers. These experiments further illustrate that DNA/RNA-binding or processing proteins are convenient targets for nanopore experiments, as those nucleoprotein complexes can be 'grabbed' by the negative charge of the nucleic acid. The Gundlach lab appropriately termed this elegant nanopore force spectroscopy 'SPRNT', short for Single-molecule Picometer Resolution Nanopore Tweezers [100,101]. Regarding noise and resolution of DNA stepping, it benefits greatly from the extremely short distance between the speed-controlling helicase and the sensing pore constriction, in contrast to e.g. optical tweezers that involve micron-long handles to the distant micron-sized beads. Note that, unlike nucleic acids, peptides lack a uniform charge distribution, which is one reason why protein sequencing is a much bigger challenge for nanopore technology (see below).

Next, we consider a DNA-free all-protein system: dihydrofolate reductase (DHFR) studied with a ClyA protein nanopore (**Fig 4B**). The Maglia lab trapped such single protein complexes for tens of seconds to interrogate their functional dynamics [98]. A positively charged tail was attached to the DHFR to enhance (electro-osmotic with electrophoretic) trapping under negative voltage, and to orient the protein in a preferential direction [95,102]. In this way, tens to hundreds of dihydrofolate to tetrahydrofolate turnovers could be sensed at one DHFR molecule. And a total of five functional states were distinguished and deciphered from the label-free current recordings, simply by systematic substrate variation: ± folate, ± NADPH, etc. The result is the real-time observation of a single DHFR protein progressing through its functional cycle involving reversible and irreversible transitions that can now be distinguished. It is this kinetic connectivity between states — that is inaccessible from ensemble kinetics including time-correlation analysis – which is the key to revealing energetically driven functional processes. Interestingly, these single-molecule measurements revealed among other things that DHFR undergoes second-long catalytic pausing, related to an off-pathway that was linked to the tolerance of high NADP+ concentrations. The Achilles heel of this elegant label-free trapping approach lies in its applicability to small proteins only, set by the pre-determined size of the (already relatively large) protein nanopore lumen (≤7nm diameter). To turn single protein trapping into a more generally applicable protein sensing tool, we recently developed the NEOtrap, the 'Nanopore Electro-Osmotic trap', by combining DNA origami and passivated solid-state nanopores that can be obtained at any size suitable for the target protein [103]. In brief, the origami structure is used to create strong electro-osmotic flows in a solid-state nanopore, which allows us to catch a protein, label-free, and hold it for several minutes at the most sensitive region of the nanopore. Alternatively, plasmonic trapping is being developed as an entirely optical way to trap and study single proteins [104,105]. Finally, many more approaches are possible. For example, bulk enzyme activity such as ubiquitination, was studied using nanopores by monitoring the time-dependent accumulation of product molecules [106]. And the kinetics of amyloid fibrillization was also observed in label-free solid-state nanopore experiments [49].

**Transient protein-protein interactions**

Specific protein-protein interactions (PPIs) have been studied at the single-molecule level using functionalized protein nanopores. For example, Movileanu and co-workers turned the FhuA β-barrel pore into a PPI sensor by fusing a Barnase domain plus an additional sensor peptide to the FhuA pore (**Fig. 5A**). Upon Barnase-Barstar interaction, the sensor peptide is pulled out of the pore, leading to a (at first counter-intuitive) higher conductance level of the bound state compared to the unbound state. In this way, the authors detected transient binding that is hardly accessible in bulk, and the specificity of this nanopore sensor was also shown in mammalian serum [107]. A similar strategy was applied earlier to detect kinase-substrate interactions and their kinetics [108,109]. Alternatively, protein pores have been decorated with aptamers, and other functional motifs turning them into specific sensors. Furthermore, protein interaction has been studied using DNA carriers and solid-state nanopores: e.g. directly for dCas9 interactions [110,111], or using epitope [112] or aptamer [113] functionalized DNA constructs as designed by the Keyser and Edel labs to specifically detect antibodies, avidin, etc. Lastly, also small molecules like sugars and single amino acids were detected by specific binders like glucose-binding protein that change their conformation upon ligand binding. Trace amounts thereof were even quantified in bodily fluids, including sweat, blood, saliva, and urine [114].

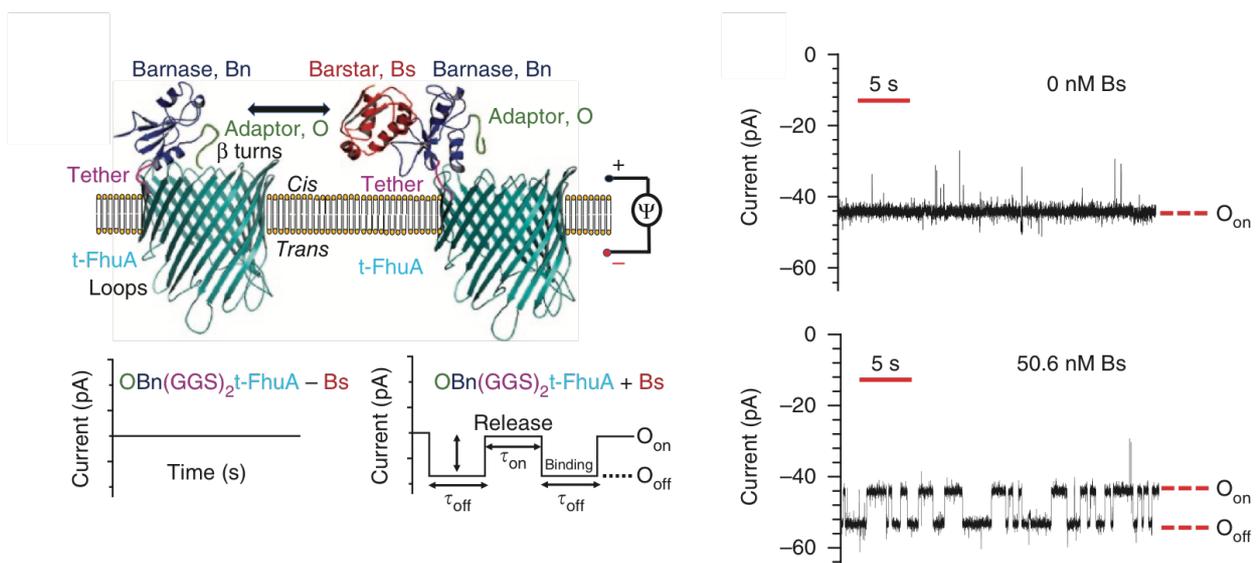

*Figure 5: Single protein-protein interactions observed with nanopores. Barnase-Barstar interactions recorded using a FhuA fusion construct that serves as a specifically functionalized protein pore sensor [115]. Two-state kinetics appear upon Barstar addition due to transient binding events, leading to a detached sensor peptide (Adaptor O) and thus higher conductance than in the unbound state.*

**A glance at protein sequencing**

As a future perspective for nanopore technology, protein sequencing is now emerging as a new hot topic [116]. Clearly, a single-molecule protein-sequencing device comparable to nanopore DNA sequencers would have a disruptive effect in the molecular life sciences. Yet, this nascent field has to deal with impressive challenges: no uniform charge along the peptide, twenty diverse building blocks (amino acids), robust folds, etc. Nevertheless, many creative routes are currently being pursued, and specifically reviewed in [116–118]. Here, we briefly sketch a few current strategies, which fall into two main categories: *de novo* sequencing and fingerprinting. The latter aims at protein identification based on prior knowledge, which is e.g. pursued by Meller and coworkers who couple solid-state nanopores as a delivery system to confocal fluorescent detection [119]. Fluorescent tags are then attached to specific amino acids in unfolded peptides, in an attempt to read their number and sequence order. Brinkerhoff and Dekker exploit Akeson and Gundlach's successful DNA-sequencing strategy described above (**Figure 4A**) and hook up a short peptide to the ssDNA strand, which is then slowly ratcheted through the MspA protein pore, with the potential to enable *de novo* sequencing of short peptides [120]. An encouraging proof-of-concept study by Oukhaled, Aksimentiev, Behrends and co-workers showed that under special conditions, a majority of all twenty amino acids could be distinguished using the aerolysin protein pore [121]. Early reports [122,123] of the unfolding and ratcheting of multi-domain proteins through a α-hemolysin protein pore by the proteasome ClpXP complex is inspiring strategies for long peptide reads. Such ideas are pursued by several groups, where also other unfolding motors are considered, e.g. SecYEG [124,125]. Besides sequencing or identification through fingerprinting, direct detection of post-translational modifications with single-molecule resolution would be another very important progress in biotechnology. Under specific conditions, phosphorylations and glycosylations were already detected in a lable-free way [106,126,127], thereby paving the way for more general implementations.

**Single-molecule techniques side by side**

How do nanopores compare to other single-molecule techniques? In **Figure 6,** we compare various popular single-molecule techniques that can sense the time evolution of a single protein: high-speed AFM, optical and magnetic tweezers, surface immobilized smFRET, and nanopores. All of these techniques can detect kinetic connectivity, i.e., distinguish a transition from state A to state B, from the inverse transition B to A. This is a crucial difference to other time-resolved techniques that detect only direction-less fluctuations and mean life-times of states, such as FCS. This kinetic connectivity is absolutely necessary if the goal is to understand protein systems with more than two kinetic states, and in particular non-equilibrium steady-states. We further see from **Figure 6** that protein dynamics occur on a broad range of timescales, where the ultimately rate-limiting steps in protein function depends on underlying processes that can be orders of magnitudes faster, which is sometimes called the hierarchy of dynamics [128]. A broad time bandwidth is therefore key to uncover the origin of protein function, energetically and quantitatively. This is exactly where nanopores excel with their broad electrical bandwidth, and new experimental strategies that permit to effectively use a large part of it to generate information. Note that the actually informative bandwidth as shown in published work (white boxes in **Figure 6)** is narrower than the purely technical bandwidth (white lines) for all techniques. Depending on the technique, it is in practice limited by drift, protein stability (affected by surface interactions, force application, labelling, reactive oxygen species), photo-bleaching and other photo-physics, or baseline stability. Altogether, nanopores stand out in their ability to detect the broad-range dynamics of proteins, which notably is achieved in solution, without the need of surface immobilization nor artificial labelling, and using affordable instrumentation.

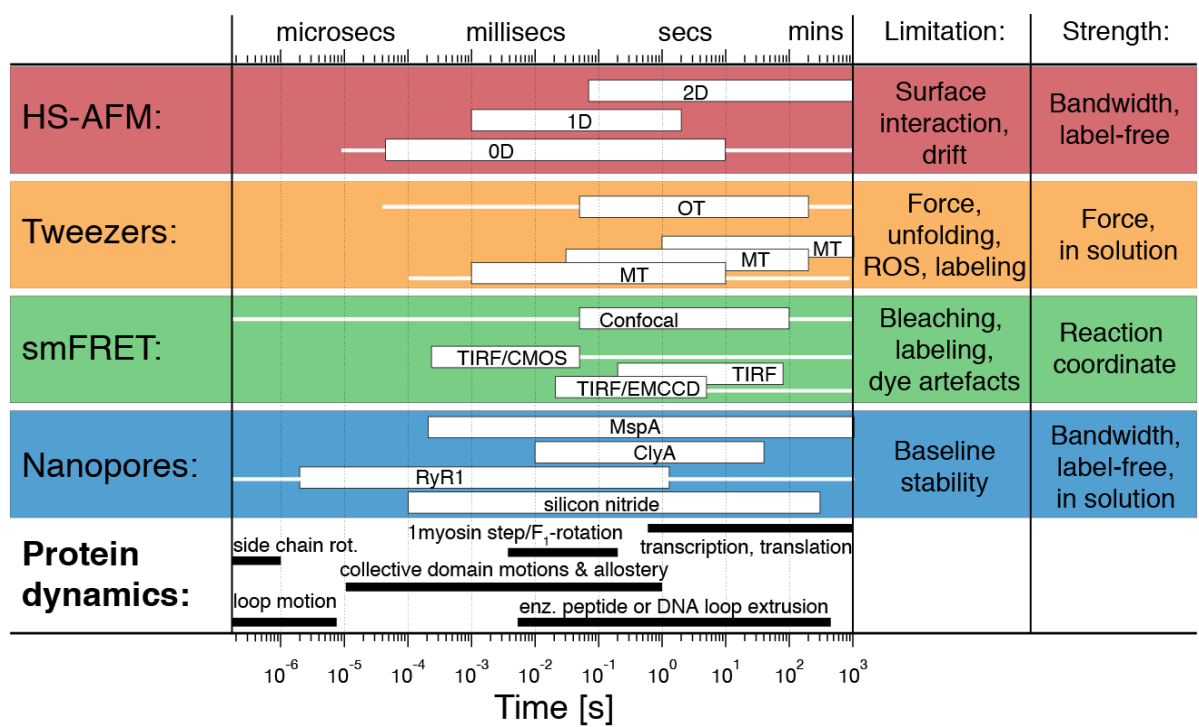

*Figure 6: Timescales of protein functional dynamics and experimentally accessible bandwidths for their observation. White boxes represent experiments reported in the literature, thin white lines indicate the technical detector bandwidth: High-speed AFM (HS-AFM) in 2D imaging mode, 1D line scans, or 0D 'on-spot' detection, as indicated [129]. Optical (OT) [130,131], and magnetic tweezers (MT) [132–136] probing protein function (not unfolding, ROS: reactive oxygen species). Surface immobilized smFRET in confocal [137], or TIRF mode with CMOS [138] or EMCCD [2] detectors. Protein [98,139,140] and solid-state nanopores (own work). Examples of protein dynamics [141–143] controlling function [5,131,144] occurring in the microsecond to minutes range. Faster dynamics, as well as ensemble studies and mechanical unfolding experiments are not considered in this figure.*

**CONCLUSION**

In this brief review, we introduced nanopore experiments, and discussed their great potential for protein science based on specific examples demonstrating the unique benefits of this technique, including (i) the label-free detection of protein function and protein interactions, (ii) at the single-molecule level, (iii) time-resolved with a bandwidth spanning currently up to seven orders in time in a single experiment, and (iv) all of this achieved with relatively cheap instrumentation compared to other single-molecule techniques. For DNA- or RNA-binding proteins, nanopore force-spectroscopy is a most convenient label-free technique with sub-basepair resolution for motor proteins stepping along nucleic acid strands. All-protein systems can be studied as well, e.g. by electro-osmotic trapping in protein or solid-state nanopores. Although these label-free experiments lack a predetermined spatial reaction coordinate (set by labelling in other techniques) dissecting individual functional states can be achieved by systematic experiments. Furthermore, transient interactions can be detected label-free and with high specificity, even in bodily fluids. With all these techniques, protein scientists are well equipped to reach beyond the static structure era, and elucidate the dynamic character of these vital biomolecules. Furthermore, a lot can be expected from the emerging community that works towards protein sequencing, where biochemists and biophysicists meet with solid-state physicists, theoreticians, organic chemists, biomedical and nano-engineers to create a flourishing environment for ground-breaking interdisciplinary research.

## SUMMARY

- **Nanopore detection is an affordable, label-free, single-molecule technique that is independent of photo-bleaching, surface immobilization, or mechanical unfolding of proteins.**

- **Long-term observation of single proteins can be achieved by electro-osmotic trapping, DNA association, or specific protein interactions.**

- **Protein functional states can be directly recognized in label-free current recordings.**

- **Electrical nanopore detection provides access to a broad time bandwidth of µs to hours matching the protein functional time range.**

- **Thousands of protein functional cycles can be observed on one single molecule at sub-millisecond resolution**.

| GLOSSARY | |
|---|---|
| 1D, 2D, 3D | 1-, 2-, 3-dimensional |
| ATP | Adenosine triphosphate |
| cryoEM | cryo electron microscopy |
| DHFR | dihydrofolate reductase |
| DNA | desoxyribonucleic acid |
| EPR | electron paramagnetic resonance |
| FCS | fluorescence correlation spectroscopy |
| FRET | Förster resonance energy transfer |
| NMR | nuclear magnetic resonance |
| RNA | ribonucleic acid |
| SANS | small angel neutron scattering |
| SAXS | small angle x-ray scattering |
| smFRET | single-molecule FRET |
| SEM | scanning electron microscope |
| TEM | transmission electron microscope |
| TIRF | total internal reflection fluorescence |

## CONFLICTS OF INTEREST
The authors declare that there are no competing interests associated with the manuscript.

## ACKNOWLEDGEMENTS
We thank Allard Katan and Richard Janissen for their comments on high-speed AFM and magnetic tweezers, respectively, and Sabina Caneva and Wayne Yang for comments on 2D materials.

## FUNDING
SS acknowledges the Postdoc.Mobility fellowship no. P400PB_180889 by the Swiss National Science Foundation. CD acknowledges the ERC Advanced Grant LoopingDNA (no. 883684) and The Netherlands Organization of Scientific Research (NWO/ OCW) as part of the NanoFront and Basyc programs.

## AUTHOR CONTRIBUTION
SS framed the concept of the review. Both authors wrote and discussed the manuscript.